\documentclass[]{elsarticle}

\usepackage{lineno,hyperref}
\usepackage{enumerate}
\usepackage{graphics}
\newcommand{\nocomma}{}

\newcommand{\tmem}[1]{{\em #1\/}}

\newcommand{\tmstrong}[1]{\textbf{#1}}
\newcommand{\tmtextbf}[1]{{\bfseries{#1}}}
\newcommand{\tmtextit}[1]{{\itshape{#1}}}
\newenvironment{enumeratealphacap}{\begin{enumerate}[A.] }{\end{enumerate}}
\newenvironment{enumerateroman}{\begin{enumerate}[i.] }{\end{enumerate}}
\newenvironment{enumerateromancap}{\begin{enumerate}[I.] }{\end{enumerate}}

\journal{Journal of \LaTeX\ Templates}

\begin{document}

\begin{frontmatter}

\title{The growth of dry convection in the conditionally stable troposphere: 
Non-adiabatic effects}

\author{E.A. Kherani, B.R. Tiwari and J.H.A. Sobral}
\address{
Instituto Nacional de Pesquisas Espaciais, \\
Av. dos Astronautas, 1728, Sao jose dos campos
}

\begin{abstract}
In this work, we study the growth characteristics of the convective
  instability (CI) in the dry troposphere by relaxing the
  adiabatic compressibility condition of Oberbeck-Boussinesq (OB) approach. 
  We derive a new non-adiabatic-Boussinesq (NAB) expression 
  for the modified Brunt-Vaisala frequency $(\omega_b)$, without considering 
  the adiabatic compressibility condition of OB approach.
  This NAB expression reduces to the known Oberbeck-Boussinesq (OB) expression under 
  adiabatic compressibility condition.
  The NAB expression of $\omega_b$ is found to be
  modified from its OB counterpart such that the stabilizing 
  adiabatic lapse rate in OB expression is replaced by a modified non-adiabatic lapse rate given as 
  $\left( \eta - 1 \right)$ times the auto-convective lapse rate. 
  Here $\eta$ is the ratio of hydrostatic density to the total density. 
  We perform numerical experiments of CI for the conditionally stable troposphere 
  i.e for the troposphere that has the environmental lapse rate negative but smaller than the adiabatic lapse rate.
  A novel feature of the present study is that the CI grows under proposed NAB approach 
  in spite the conditionally stable condition and remains suppressed under OB approach.
  The present study, thus,
  proposes an alternative NAB approach for the positive growth of the CI in the 
  dry troposphere for which the CI is conditionally stable under OB approach.
\end{abstract}

\begin{keyword}
Tropospheric convection, Non-adiabatic flows, Numerical simulation
\end{keyword}

\end{frontmatter}


\section{Introduction}

The convection in the troposphere of the terrestrial atmosphere is a natural
phenomenon, that determines the meteorological conditions such as convective
storms, thunderstorms, small and large-scale circulations etc. The convection
arises owing to the convective instability (CI) in the presence of
destabilizing negative environmental temperature gradient that
competes with the stabilizing adiabatic temperature gradient \cite{1,2,3,4}. 
These gradients are respectively, characterized by the environmental
lapse rate $\gamma_e = \frac{1}{T} \frac{\partial T}{\partial y}$ and the
adiabatic lapse rate $\gamma_{_{ \rm{ad}}} = \frac{g}{c_p T}$.

Theoretical works concerning the tropospheric CI are in abundance and most of
them are based on the Oberbeck-Boussinesq (OB) approach. The four most
important assumptions under OB approach are the following \cite{3,5}:
\begin{enumerateroman}
  \item The shallow atmosphere condition for the ambient hydrostatic
  atmosphere which means that the density scale
  height is approximately equal to the pressure scale height,
  
  \item Isothermal incompressible flow (time restrictions on the
  perturbation such that the buoyancy velocity remains slower than 
  the acoustic velocity),
  
  \item A linearization of $\frac{1}{\rho}$ where $\rho$ is the total density,
  consists of hydrostatic equilibrium density and the linearly perturbed density,
  
  \item Adiabatic compressible flow.
\end{enumerateroman}

These assumptions simplify the instability treatment and a linear
Brunt-Vaisala frequency (or growth rate) expression for the
Oberbeck-Boussinesq-Convective-Instability (OBCI) is obtained in the following
form \cite{3}:
\[ \omega_{ \rm{bB}} = \sqrt{g \left( \gamma_{ \rm{ad}} + \gamma_e \right)} \]

In this conservative form, the OBCI may not be excited since usually $|\gamma_e|
< \gamma_{ \rm{ad}}$ for the radiative-convective equilibrium temperature
profile in the troposphere \cite{6} 
i.e, the troposphere is conditionally (or adiabatically) stable to the OBCI. 
Even during maximum
surface heating conditions, $|\gamma_e|$ becomes larger than
$\gamma_{ \rm{ad}}$ only within a part of the troposphere but not in the
entire troposphere. In order to account for the observed convective dynamics,
instead of $\gamma_{ \rm{ad}}$, the average adiabatic lapse rate
$\gamma^{ \rm{av}}_{ \rm{ad}} = 6.5 / T$ or the moist or wet adiabatic lapse
rate $\gamma^m_{ \rm{ad}} < \gamma^{ \rm{av}}_{ \rm{ad}}$ are considered
\cite{7} and CI simulation are performed for the moist convection that
do not satisfy some or all of the OB assumptions \cite{8,9,10}.
Also, aforementioned OB assumptions may not hold during non-linear phase of 
CI and numerous studies have been
performed for the CI in the dry
troposphere to account for the the non-linear
dynamics \cite{12,13,14,15,16}.


Lilly \cite{17} has derived the governing equations for the OBCI 
incorporating non-adiabatic dynamics i.e. without the fourth OB assumption. 
These equations are employed 
to study the non-adiabatic effects for conditionally unstable troposphere.
However, to date, the studies concerning the non-adiabatic contributions
arising from the non-adiabatic nature of compression are not pursued for the conditionally stable dry troposphere. 
It should be noted that the conditionally stable condition $(|\gamma_e| < \gamma_{ \rm{ad}}, \gamma_e < 0)$ 
that prevails in the most part of the troposphere,
invalidates the fourth OB assumption since the temperature gradient is not 
strong enough to support the adiabatic compression which is a rapid process. 
Thus, the non-adiabatic contributions are expected during linear
phase of the CI for conditionally stable troposphere and 
this aspect is not investigated so far. 
Moreover, an analytical expression for the
Brunt-Vaisala frequency or growth rate under non-adiabatic compressibility condition is always desirable 
and it is not yet available. 
These aspects are the focus of the present study. 
Hereby, we define a non-adiabatic-Boussinesq (NAB) approach that consider all OB assumptions 
except the last assumption i.e. adiabatic compressibility condition. 
The corresponding convective instability is referred as non-adiabatic-Boussinesq-CI (NABCI).
The objectives of the present
study are the following:
\begin{enumerateromancap}
  \item Derivation of an analytical expression for $\omega_b$ of NABCI having a form
  similar to $\omega_{ \rm{bB}}$,
  
  \item To search for the possibilities of positive growth of CI under condition 
  $(|\gamma_e| < \gamma_{ \rm{ad}}, \gamma_e < 0)$ i.e., under stable OBCI condition,
  
\end{enumerateromancap}

With these objectives, we develop a theoretical framework for the CI in the
dry troposphere in appendices A-C. In appendix A, we derive the governing
equations for the CI using inviscid hydrodynamic equations. These governing
equations include the growth (or wave) equation for the vertical wind $w_y$,
temperature equation and the perturbed density (total density minus
hydrostatic density) equation. The growth equation contains the general
expression of Brunt-Vaisala frequency $\omega_b$. In appendices B and C 
respectively, the $\omega_b$ expression is derived without the adiabatic compressibility
condition (non-adiabatic-Boussinesq or NAB approach) 
and with all four OB assumptions (OB approach).
Below, in section 2.1, we analyze the derived equations qualitatively. In
section 2.2, a one dimensional (in the vertical direction representing the
altitude) simulation of the CI is presented to study the NAB contributions for
chosen initially adiabatically stable temperature profile.

\section{Results and Discussion}

\subsection{The growth of CI under proposed NAB approach: a qualitative analysis}

In the appendix B, we have derived a non-adiabatic expression (B9) of the
Brunt-Vaisala frequency in the following form:
\begin{eqnarray}
  \omega_{b^{}}^2 = \eta g \left( \gamma_d + \gamma_e \right)  &  \quad \rm{where} \quad &
  \gamma_d = \left( \eta - 1 \right) \gamma_{ \rm{ac}}  \quad \rm{and} \quad \eta =
  \frac{\rho_h}{\rho} \approx 1-\frac{\rho_t}{\rho_h}
\end{eqnarray}
where $\gamma_d$ is defined as the non-adiabatic lapse rate, 
$(\rho_h,\rho_t,\rho)$ are the hydrostatic density, perturbed density and 
the total density. 
Equation (1) is obtained under
shallow atmosphere, isothermal incompressible and linearization conditions which are the
first three OB assumptions. 
Remaining OB assumption, i.e. adiabatic
compressibility is not employed and thus this
expression is a non-adiabatic expression of $\omega_b$ representing the
NAB approach.

$\eta$ in (1) is the measure of degree of density change within a heated
fluid parcel inside which $\rho \leq \rho_h \Rightarrow \eta \geq
1$. Thus, the non-adiabatic lapse rate $\gamma_d$ is always positive within a
heated fluid parcel i.e. this term has the stabilizing effects. 
For example, $\eta = \left(
1.01, 1.3 \nocomma, 2, 10 \right) $ correspond to the degree of density change
equals to (1\%,20\%,50\%,90\%) respectively representing the initiation,
linear, weakly non-linear and highly non-linear phases of the convection,
respectively. The expression (1) is, however, valid only in the
linear phase of convection when the shallow convection condition is
satisfied.

In the appendix C, we have employed the last OB assumption (i.e. the adiabatic compressibility condition) 
and found that $\eta = \gamma$, where $\gamma$ is the ratio of specific heats. 
Thus the adiabatic compressibility assumption in OB
imposes the condition on the degree of density change such that it should be
approximately equals to $\frac{\gamma-1}{\gamma}\sim 30$\% which can be taken as the upper limit for
the linear phase of the convection. 
The question is what happens to the CI from beginning, when the degree of 
density change is insignificant, up-to the upper limit of the linear phase
i.e., during the phase when $\eta$ increases from 1 to $\gamma$.
It is evident from (1) that as $\eta$
increases from 1 to $\gamma$, the stabilizing term also increases.
Therefore, the last OB assumption i.e. the adiabatic compressibility enforces
the stabilizing effects. It is thus expected that from the initiation to the
upper limit of linear phase of convection i.e from the NAB phase to the OB
phase, the required threshold $\gamma_e$ for the instability increases from
$\sim 0.01 \gamma_{ \rm{ac}} \ll \gamma_{ \rm{ad}}  \rm{to} \left( \gamma - 1
\right) \gamma_{ \rm{ac}}=\gamma_{ad}$ as shown in appendix C. 
It is thus evident that during
the NAB phase, the convection convincingly grows in the troposphere since the
stabilizing non-adiabatic lapse rate is much smaller than the adiabatic lapse
rate.

It should be pointed out that the present study deals with the dry convection
and thus the non-adiabatic effects discussed above are not owing to the presence of
moist or another phase of matter than the gas. The non-adiabatic effects, in the
present study, are owing to the linear phase $\left( \eta \sim 1 \right)$ of
the convection which is not fast enough (owing to $\gamma_e <
\gamma_{ \rm{ad}}$ for the radiative-convective equilibrium temperature
profile) to undergo the spontaneous or adiabatic compression.

\subsection{The growth of CI under proposed NAB approach: a quantitative analysis}

To study aforementioned aspects quantitatively, we develop a simulation model
of the CI where the following set of equations are simultaneously solved:
\[
  \frac{\partial^2 w_y}{\partial t^2} = - \omega^2_b w_y - \eta g \frac{\partial
  w_y}{\partial y} 
\]
\[
  \frac{\partial \rho_t}{\partial t} = w_y  \left( \rho_h + \rho_t \right)
  \gamma_e,  \quad \quad
  \frac{\partial \rho_h}{\partial t} = - w_y  \frac{\partial \rho_h}{\partial
  y}
\]
\begin{equation}  \frac{\partial T}{\partial t} = - w_y T \gamma_e,
\quad \quad 
\gamma_e=\frac{1}{T}\frac{\partial T}{\partial y}
\end{equation}

Here, $\left( w_y \nocomma, \rho_t \nocomma, \rho_h, T \right)$ are the wind,
perturbed density inside the heated fluid parcel, the hydrostatic density and
the total temperature respectively. The first equation in (2) is the equation (A3)
in the appendix A. 
Using continuity equation (A1) and equation (B8) in appendices A-B, two
simultaneous equations for $\left( \rho_t, \rho_h \right)$ are obtained in
(2). 
The temperature equation in (2) is obtained using the pressure and continuity equations 
in (A1).
\\
It should be mentioned that the incompressibility condition 
decouples the mechanical and thermodynamical processes \cite{17} and leads to 
the continuity equations for $(\rho)$ without the non-adiabatic term and 
the continuity equation for $(\rho_t)$ with the non-adiabatic term, 
as derived by Lilly \cite{17} as equations (2.2), (2.9) and (2.11).
In the present study, the derived continuity equations for $\rho$ 
in equation (A1) is identical to equation (2.2, 2.9) derived by Lilly \cite{17}.
Moreover, the continuity equation for $\rho_t$ as derived above in (2) is similar to 
the equation (2.11) of Lilly \cite{17} such that it retains the non-adiabatic 
contribution through $\gamma_e$ term. 
\\
\\
The set of equations in (2) are solved using FTCS finite-difference,
Crank-Nicolson implicit integration scheme and \ Successive-Over-Relaxation
method. This numerical scheme is discussed in detail by \cite{18,19} 
to solve the convective instability in the ionospheric plasma and to
solve the acoustic-gravity wave equation in the atmosphere. The non-local term
in the growth equation of $w_y$ is not considered in the present study since
our objective is to understand the non-adiabatic effects that is owing to the
non-adiabatic lapse rate $\gamma_d$ in $\omega^2_b$ term.

The simulation domain is one-dimensional, consists of altitude covering 0-20
km altitude with grid resolution $\Delta y = 0.5  \rm{km}$. The initial
hydrostatic density $\rho_h$ profile is considered to be the exponential
decreasing with height with scale height equals to $20  \rm{km}$. In Figure
1a, the chosen radiative-convective equilibrium temperature profile is shown.
In Figure 1b, the initial environmental lapse rate $\gamma_e$ multiplied by
the initial temperature $\left( T_o \right)$, is plotted. For comparison,
the $\gamma_{ \rm{ad}} ~ \rm{and}~\gamma_{ \rm{ac}}$ multiplied
by $T_o$ are also plotted. It is evident from figure 1b that the chosen temperature profile
satisfies $|\gamma_e| < \gamma_{ \rm{ad}}$ i.e the profile is adiabatically
stable in the entire troposphere.

We present the simulation results by performing three numerical experiments
corresponding to three approaches which are the NAB, OB and a general approach.
These three approaches correspond to three different expression for $\omega^2_b$ 
in (2) and is respectively given by expressions (1), (C1) and (A4) as follows:

\begin{eqnarray}
  \omega_{b^{}}^2 = \eta g \left[ \left( \eta - 1 \right) \gamma_{ \rm{ac}} +
  \gamma_e \right]  ~\rm{where} ~\eta =\frac{\rho_h}{\rho} \approx 1- \frac{\rho_t}{\rho_h}  \rm{(NAB ~ approach)}
\\ \nonumber
  \omega_{bB}^2 = g \left( \gamma_{ \rm{ad}} + \gamma_e \right)  \rm{(OB~approach)}
\\ \nonumber
  \omega^2_{bg} = \frac{1}{\rho^2} \frac{\partial p}{\partial y} \frac{\partial
  \rho}{\partial y} - \frac{1}{\rho} \frac{\partial^2 p}{\partial y^2}
   \rm{(a~general~approach)}
\end{eqnarray}

In this section, we present the results corresponding to NAB approach while 
in the next section, the results from other two approaches are presented 
and compared with the results from the NAB approach.

The source of perturbation can be either of wind type or thermal type. We
study a wind type perturbation such that a uniform (over
altitude) amplitude $= 10^{- 3} m / s$ at t=0 is chosen.
With this perturbation, the temporal-altitude evolution of the
$\left( w_y, \rho_t \nocomma, \rho_h  ~\rm{and} ~T \right)$ are simulated. In
Figures 2, the results are presented. In Figures 2a-2c respectively, 
the temporal-altitude
evolution of $\eta = \frac{\rho_h}{\rho_h + \rho_t} \nocomma, ~T \gamma_e =
\frac{\partial T}{\partial y}$  and $w_y$ are shown. In 2d, the
temporal evolution of the maximum (over altitude) value $\left( w_y^{\max}
\right)$ of $w_y$ is plotted. We may note the following characteristics from Figure 2:
\begin{enumeratealphacap}
  \item $w_y^{\max}$ grows linearly and then exponentially for first 10
  minutes and then becomes constant,
  
  \item With time, ($\eta$, $T$ and $w_y$ ) grow in
  amplitudes in the altitude region where $\gamma_e < 0$ initially and they
  attain the largest amplitudes near the altitude where the initial
  $\gamma^{}_e$ has maximum negative value,
  
  \item At $\eta = \gamma$ that occurs at t=10 minutes, $|\gamma_e|$ becomes
  equal to $\gamma_{ \rm{ad}}$ and $w_{y}^{\max}$ stops growing and becomes
  saturated to the value of 20 m/s.
  
  \item After this time, though, both  $\eta ~\rm{and}~\gamma_e$ continue to
  grow. \ 
\end{enumeratealphacap}
The exponential growth of $w_y^{\max}$ under characteristic (A) is an
indication of the linear phase of the CI. The growth suggests that the small
wind perturbation of $\sim 10^{- 3} m / s$ has grown to 20 m/s within 10
minutes under the action of CI. The characteristics under (A-B) suggest that
in spite of the initial adiabatic stable temperature profile, the CI grows in
the troposphere. The growth is owing to the non-adiabatic nature of the CI during
linear or NAB phase $\left( \eta < \gamma \right)$ when the stabilizing lapse
rate $\left( \eta - 1 \right) \gamma_{ \rm{ac}}$ remains smaller than the
$|\gamma_e|$. As time progress, both $\eta  ~\rm{and} ~\gamma_e$ and thus both
stabilizing and destabilizing lapse rates grow, leading to the saturation, as
indicated by characteristic (C).

The characteristic under (C) also suggests that as the tropospheric state approach
to the adiabatic state of gas, $|\gamma_e|$ approaches $\gamma_{ \rm{ad}}$.
This suggests the consistent adjustment of the temperature profile following
the nature of compression within the heated fluid parcel 
and consistent transition of CI from NAB to OB type. As $|\gamma_e|$
approaches $\gamma_{ \rm{ad}}$, the CI begins to slow down but 
the growth rate remains positive, as suggested by characteristic (D). 
\ \ \

It is evident from Figure 2(b) that $|\gamma_e| < \gamma_{ \rm{ac}}$ is
maintained in the troposphere. Thus, the shallow convection condition, which
is employed to derive the expression (1) is always satisfied, during the
simulation. The present simulation is not applicable for the non-linear phase
$\left( \eta > 2 \right)$ of the convection since the shallow convection
condition may not be satisfied under non-linear phase. 
Figures 2(a-b) suggest that during the simulation, the linearity and 
the shallow atmosphere conditions are maintained i.e. the ambient scales 
satisfy the condition of linear-first order approximation. 
This classifies the proposed NAB approach as of Boussinesq in nature. 
\ \

\subsection{The proposed NAB approach vs. OB and a general approach}
In this section, we present the results from three numerical experiments
corresponding to three approaches which are the NAB, OB and the general approach.
As mentioned in the last section, these three approaches are different in terms 
of $\omega^2_b$ expression which are respectively given in (3).
\\
In Figure 3, the time evolution of $w^{max}_y$ is shown under three numerical experiments.
The evolution under NAB approach, as shown in Figure 2d, is replotted as a blue curve in 
Figure 3. We may note that under OB and the general approaches, $w^{max}_y$
shows periodic variation in time such that the mean amplitude is decreasing with time. 
Under OB approach, the periodic variation is expected since the ambient atmosphere is
adiabatically stable leading to the positive $\omega^2_{bB}$ in expression (3).
Under general approach, similar oscillating behavior as under OB approach suggests that 
instability is not excited when none of the four OB assumptions are implemented. 
In contrast to OB and general approaches, the proposed NAB approach with the new expression 
for $\omega^2_b$, reveals a exponential growth, as discussed in the last section.
Therefore, the proposed mechanism assists the growth of the CI for the troposphere 
which is stable under OB and the general approaches.

The possible reason for different growth characteristics under NAB and OB approaches is as follows: 
Under OB approach, that imposes the adiabatic (or rapid) condition, 
the instability does not have sufficient time to extract the free energy from the 
negative temperature gradient. 
On contrary, under proposed NAB approach, the instability acquires sufficient time 
to extract energy since no adiabatic or rapid condition is imposed.

A new set of governing equations of the CI, given by equation (2), provides
an alternative mechanism for the positive growth of the instability in the
adiabatically stable troposphere. These set of equations do not incorporate
moist convective dynamics, but rather retains the non-adiabatic effects of the dry
convection. The non-adiabatic effects, here, are arising owing to the small
amplitude of density perturbation $\eta < \gamma$ in the presence of weak
$|\gamma_e| < \gamma_{ \rm{ad}}$. The convection of two types, i.e. NAB 
 and OB can be studied with the identical set of equations, as shown in the
present study. In future, we intend to extend this simulation study for the
two and three dimensional convective instability, whose results may directly
be compared with the observations.

\section{Summary}

In this work, we present an alternative Non-adiabatic Boussinesq (NAB) approach 
of the convective instability (CI) without imposing the adiabatic compressibility condition 
in the dry troposphere. 
Lilly \cite{17} had presented the NAB framework for the CI which is often applied 
for the conditionally unstable dry and moist troposphere. 
However, for the conditionally stable dry troposphere, such studies are not pursued and is the focus of
the present study. 
In the conditionally stable dry troposphere, non-adiabatic effects may arise since 
the temperature gradient or
the environmental lapse rate is weaker than the adiabatic lapse rate and may 
not sustain the rapid dynamics arising owing to the adiabatic compressibility condition 
of OB approach.
\\
\\
We have derived a new Brunt-Vaisala frequency $\left( \omega_b \right)$ expression
for the instability of NAB type. 
Under proposed NAB approach, $\omega_b$ is
found to be modified from its OB counterpart such that the stabilizing
adiabatic lapse rate $\gamma_{ \rm{ad}}$ is replaced by a modified lapse rate
$\left( \eta - 1 \right) \gamma_{ \rm{ac}}$ . Here, $(\gamma_{ \rm{ac}}
\nocomma \nocomma, \eta)$ are the auto-convective lapse rate and the ratio of
hydrostatic density to the total density. 
It is shown that with the adiabatic compressibility condition, $\eta$ becomes equal to 
the specific heat ratio ($\gamma$) and the new NAB expression of 
$\omega_b$ reduces to the known OB expression.

Therefore, the adiabatic compressibility condition under OB approach imposes the condition 
on $\eta$ which, in turn, imposes the condition on the degree of density change to be approximately equals 
to 30$\%$ (for $\gamma \sim 1.4-1.6$) of the hydrostatic density. 
The question is what happens to the CI during $0\%-\sim 30$\% of degree of density change, 
i.e., during the initial phase when $\eta$ increases from 1 to $\gamma$ ?
In this work, we seek to answer this question by proposing the NAB approach with 
a new expression for $\omega_b$ that retains 
the $\eta$ without imposing compressibility condition on it 
and allows the instability to evolve 
self-consistently from $\eta \approx 1$ to $\eta=\gamma$ i.e from NAB to OB.

We further obtain the governing equations for the wind, temperature and density 
and perform numerical experiment under NAB approach for the conditionally stable dry troposphere.
The novel features of these experiments are as follows: 
(a) In spite the initial conditional stable temperature profile, the CI grows exponentially in 
the troposphere owing to
the non-adiabatic nature of the CI during linear phase $\left( 1< \eta < \gamma
\right)$, (b) As the tropospheric state approach to the adiabatic state of
gas i.e. $\eta \rightarrow \gamma$, 
$|\gamma_e|$ approaches towards $\gamma_{ \rm{ad}}$ 
suggesting the consistent adjustment of the temperature profile and 
consistent transition of CI from NAB to OB type.

The simulation is also carried out for two other approaches which use 
a general and OB expressions of $\omega_b$. In contrast to the exponential growth under proposed NAB approach, 
the general and OB approaches reveal periodic variation in time 
suggesting the stability of CI. 
In the absence of adiabatic compressibility condition, the CI acquires sufficient time to extract the 
energy from the unstable negative temperature gradient and subsequently grows under NAB approach. 
Therefore, the proposed NAB mechanism assists the growth of the CI for the troposphere 
which, otherwise, is stable under OB and the general approaches.

\section{Acknowledgement}
First author E.A.K. wish to thank financial support from 
Fundac\~ao de Amparo \`a Pesquisa do Estado de S\~ao Paulo (FAPEPSP) under process 2011/21903-3.

\newpage

\appendix

\section{Governing growth equation of the instability}

We consider the atmospheric density $\left( \rho \right)$ and pressure $\left(
p \right)$ to have small variation in the horizontal direction $\left( \hat{x}
\right)$, in comparison to the large variation in the vertical direction
$\left( \hat{y} \right)$ so that we may neglect the horizontal fluid advection
in comparison to the vertical fluid convection. On the other hand, the
vertical $\left( \hat{y} \right)$ and horizontal $\left( \hat{x} \right)$
variations of the wind $\left( \vec{w} \right)$ is governed by the
incompressible flow condition. Under such
horizontally-stratified-incompressible conditions, the hydrodynamic equations
may be written in the following form:

\begin{eqnarray}
  \frac{\partial \rho}{\partial t} = - w_y  \frac{\partial \rho}{\partial y} &
  ; & \frac{\partial p}{\partial t} = - w_y  \frac{\partial p}{\partial y}; \hspace{1em} p = \rho RT 
\end{eqnarray}
\begin{eqnarray}
  \frac{\partial w_y}{\partial t} = - \frac{1}{\rho}  \frac{\partial
  p}{\partial y} - g & ; &  \frac{\partial w_y}{\partial y} + \frac{\partial
  w_x}{\partial x} = 0 
\end{eqnarray}

Taking a time derivative of the momentum equation (A2) and substituting
expressions from (A1), leads to the following equation for the vertical wind
$w_y$:
\[ \frac{\partial^2 w_y}{\partial t^2} = \frac{1}{\rho}  \left[ w_y 
   \frac{\partial^2 p}{\partial y^2} + \frac{\partial p}{\partial y} 
   \frac{\partial w_y}{\partial y} \right] - \frac{1}{\rho^2}  \frac{\partial
   p}{\partial y} \left[ \frac{\partial \rho}{\partial y} \right] w_y \]
or
\begin{equation}
  \frac{\partial^2 w_y}{\partial t^2} = - \omega^2_b w_y + \frac{1}{\rho} 
  \frac{\partial p}{\partial y}  \frac{\partial w_y}{\partial y} 
\end{equation}
where {\tmem{}}$\omega_b $ is the Brunt-Vaisala frequency defined by following
expression:
\begin{equation}
  \omega^2_b = \frac{1}{\rho^2} \frac{\partial p}{\partial y} \frac{\partial
  \rho}{\partial y} - \frac{1}{\rho} \frac{\partial^2 p}{\partial y^2}
\end{equation}
Depending on whether $\omega_b$ is negative (positive), $w_y$ may grow (oscillate)
in time, leading to the instability (gravity wave). Kherani et al \cite{18,19}
have used compressible form of (A3) to study the acoustic-gravity wave
dynamics that was earlier derived by \cite{20}. The second term in (A3) is a non-local term that determines the
preferred wavelength of the growing instability or the gravity wave. 
The governing growth equation (A3) of \ similar form is derived in the
past under OB assumptions and without the non-local term \cite{21,22}. 
%
\section{Non-adiabatic-Boussinesq (NAB) expression of $\omega_b$}
In this section, we derive the expressions for density and pressure gradients 
appearing in (A4), by applying first three OB assumptions mentioned in the 
introduction. In other words, except the adiabatic compressibility condition, 
other three OB assumptions are considered.
\\
\\ 
We consider $\rho$ to be composed of the hydrostatic equilibrium density
$\rho_h$ and a non-hydrostatic (or perturbed) density $\rho_t$, i.e.,
\begin{equation}
  \rho = \rho_h + \rho_t ; \hspace{1em} ; p = p_h + p_t ; \hspace{1em} \rho_h
  = - \frac{1}{g}  \frac{\partial p_h}{\partial y} ; \hspace{1em} \rho_h +
  \rho_t = \frac{p_h + p_t}{RT}
\end{equation}
\\
\\

\begin{enumerate}
  \item Isothermal-incompressible condition (first OB assumption): The isothermal incompressibility 
  is defined as follows:
\begin{equation}
  \frac{1}{p} \frac{\partial p_t}{\partial y} = 0
\end{equation}

  \item Shallow-atmosphere condition (second OB assumption): 
  Using the state of gas equation for the hydrostatic atmosphere 
  $\rho_h=\frac{p_h}{RT}$, the hydrostatic density scale height may be written as follows:
  \[
  \frac{1}{\rho_h} \frac{\partial \rho_h}{\partial y} = \frac{1}{p_h} \frac{\partial
  p_{h}}{\partial y} - \frac{1}{T} \frac{\partial T}{\partial y}
  \]
  Substitution of the hydrostatic condition $\frac{\partial p_h}{\partial y}=-\rho_h g$ 
together with $p_h=\rho_h R T$, leads to the following expression:
  \[
  \frac{1}{\rho_h} \frac{\partial \rho_h}{\partial y} = - \gamma_{{ac}}- \gamma_e 
  \]

  Here we define $\gamma_{ac}=\frac{g}{RT}$ and $\gamma_e=\frac{1}{T}\frac{\partial T}{\partial y}$
  where $\gamma_{{ac}} \rm{and} \gamma_e$ are the auto-convection and
  environmental lapse rates respectively.

 The shallow atmosphere condition is defined as $\gamma_{{ac}} > |\gamma_e|$ 
 that leads to the following condition:
 \begin{equation}
  \frac{1}{\rho_h} \frac{\partial \rho_h}{\partial y} \approx - \gamma_{{ac}} 
 \end{equation}  

  i.e., the hydrostatic density scale height equals to 
 the hydrostatic pressure scale height
 
  \item{Expression for $\frac{1}{\rho}\frac{\partial \rho}{\partial y}$ in $\omega_b$:}
In general, using the state of gas equation $\rho=\frac{p}{RT}$, 
the expression for $\frac{1}{\rho}\frac{\partial \rho}{\partial y}$ may be written as follows:
\begin{equation}
  \frac{1}{\rho} \frac{\partial \rho}{\partial y} = \frac{1}{p} \frac{\partial
  p_{}}{\partial y} - \frac{1}{T} \frac{\partial T}{\partial y}
\end{equation}
Substituting $\rho=\rho_h+\rho_t$ and $p=p_h+p_t$ from (B1), above equation reduces to 
the following expression:
\[
  \frac{1}{\rho} \frac{\partial \rho}{\partial y} = 
  \frac{1}{p} \frac{\partial p_{h}}{\partial y}+
  \frac{1}{p} \frac{\partial p_{t}}{\partial y}
   - \frac{1}{T} \frac{\partial T}{\partial y}
\]
Substitution of the hydrostatic condition $\frac{\partial p_h}{\partial y}=-\rho_h g$ 
together with $p=\rho R T$ from (B1), leads to the following expression:
\[
  \frac{1}{\rho} \frac{\partial \rho}{\partial y} = - \frac{\rho_h}{\rho}
  \frac{g}{R T} + \frac{1}{p} \frac{\partial p_t}{\partial y} - \frac{1}{T}
  \frac{\partial T}{\partial y} 
\]

Here we define $\frac{\rho_h}{\rho}=\eta$, $\gamma_{ac}=\frac{g}{RT}$ and 
$\gamma_e=\frac{1}{T}\frac{\partial T}{\partial y}$
where $\gamma_{{ac}} \rm{and} \gamma_e$ are the auto-convection and
environmental lapse rates respectively. With these definitions, 
above expression reduces to the following form:
\[
 \frac{1}{\rho} \frac{\partial \rho}{\partial y} = 
 - \eta \gamma_{{ac}} - \gamma_e +
  \frac{1}{p} \frac{\partial p_t}{\partial y}
\] 
Imposing the isothermal-incompressible condition (B2),
following expression for $\frac{1}{\rho}\frac{\partial \rho}{\partial y}$ 
in $\omega_b$ is obtained:
\begin{equation}
  \frac{1}{\rho} \frac{\partial \rho}{\partial y} = - \left( \eta
  \gamma_{{ac}} + \gamma_e \right)
\end{equation}

  \item Expression for $\frac{1}{\rho}\frac{\partial p}{\partial y}$ in $\omega_b$: 
  Using $p=p_h+p_t$, the pressure gradient may be written as follows:
  \[\frac{\partial p}{\partial y}=\frac{\partial p_h}{\partial y}+\frac{\partial p_t}{\partial y}\]
  or
   \[\frac{1}{\rho}\frac{\partial p}{\partial y}=\frac{1}{\rho}\frac{\partial p_h}{\partial y}
   +\frac{1}{\rho}\frac{\partial p_t}{\partial y}\]
   or 
   \[\frac{1}{\rho}\frac{\partial p}{\partial y}=-\frac{\rho_h}{\rho}g
   +\frac{1}{\rho}\frac{\partial p_t}{\partial y}\]
   Imposing the isothermal incompressibility condition (B2), following 
   expression for $\frac{1}{\rho}\frac{\partial p}{\partial y}$ in $\omega_b$ may be obtained:
   \begin{equation}
     \frac{1}{\rho}\frac{\partial p}{\partial y}=-\frac{\rho_h}{\rho}g =-\eta g
   \end{equation}
   Further differentiating (B6) leads to the following expression:
   \[\frac{\partial^2 p}{\partial y^2}=-g\frac{\partial \rho_h}{\partial y}\]
    
  Using shallow atmosphere condition (B3), we obtain the following expression 
  for $\frac{1}{\rho} \frac{\partial^2 p}{\partial y^2}$ in $\omega_b$:
  \begin{eqnarray}
  \frac{1}{\rho} \frac{\partial^2 p}{\partial y^2} \approx g\frac{\rho_h}{\rho} \gamma_{ac} 
  = \eta g \gamma_{{ac}} 
  \end{eqnarray}

  \item Incompressible condition: At this point, it is important to explain 
  why condition (B2) is referred as the isothermal-incompressible condition in the present study ?
  We begin with equation
  (B5) which is obtained by imposing isothermal-incompressible condition and can be written as follows:
  \[\frac{1}{\rho} \frac{\partial (\rho_h+\rho_t)}{\partial y} = - (\eta \gamma_{ac}+\gamma_e)\]
  or
   \[\frac{1}{\rho} \frac{\partial \rho_h}{\partial y}+\frac{1}{\rho} \frac{\partial \rho_t}{\partial y} 
   = - (\eta \gamma_{ac}+\gamma_e)\]
  
  Imposing the shallow-atmosphere condition (B3), following equation is obtained:
  \[-\eta \gamma_{ac}+\frac{1}{\rho} \frac{\partial \rho_t}{\partial y} 
   = - (\eta \gamma_{ac}+\gamma_e)\]
  or
\begin{equation}
  \frac{1}{\rho} \frac{\partial \rho_t}{\partial y} = - \gamma_e
\end{equation}
which is the governing equation for the isothermal incompressibility stating that 
the density change is owing to the thermal expansion. For this reason, the condition (B2) is 
termed as the isothermal incompressible condition. 
The shallow atmosphere condition together with the isothermal
incompressible condition (B2) ensures the incompressibility in the troposphere
\cite{5}.
  
  \item NAB expression of $\omega^2_b$: Substituting (B5-B7) in (A4), we 
  obtain following expression of $\omega^2_b$:
  \begin{equation}
  \omega_{b^{}}^2 = \eta g \left[ \left( \eta - 1 \right) \gamma_{{ac}} +
  \gamma_e \right] \rm{where} \eta = \frac{\rho_h}{\rho}
  \end{equation}
  This is a general NAB expression for the Brunt-Vaisala
  frequency, obtained in the present study.
  
  \item Linearization of $\eta$ appearing with $\gamma_{ac}$:
  The third OB assumption is the linearization of $\frac{1}{\rho}$ or $\eta$ in
  (B9) in a following manner \cite{3}:
  \[ \eta = \frac{\rho_h}{\rho} = \frac{\rho_h}{\rho_h \left( 1 +
   \frac{\rho_t}{\rho_h} \right)} \approx 1 - \frac{\rho_t}{\rho_h} \]
  This linearization is done only for $\eta$ appearing with $\gamma_{{ac}}$ 
  where gravity is coming with $\gamma_{{ac}}$ i.e. with the hydrostatic
  pressure gradient force. For $\eta$ appearing only with gravity can be taken
  as unity. With these linearization scheme, (B9) reduces to the following form:
  \begin{equation}
    \omega_{b^{}}^2 = - g \left( \frac{\rho_t}{\rho_h} \gamma_{{ac}} - \gamma_e  \right)
  \end{equation} 
  
\end{enumerate}

\section{Oberbeck-Boussinesq (OB) expression of $\omega_b$}

\begin{enumerate}
  
  \item Adiabatic Compressibility condition:
  The adiabatic compressible condition, which is the last remaining assumption under OB approach,
  is described by the following law of thermodynamics:
  \[
  \frac{1}{\rho} \frac{\partial \rho}{\partial y} = \frac{c_p}{\gamma R}
  \gamma_e
  \]

  Substitution of $\gamma_e$ from the isothermal-incompressible equation (B8) reduces 
  the above equation of the adiabatic compressible equation into following condition on $\eta$ :
  \[
  \frac{\rho_t}{\rho} = - \frac{\gamma R}{c_p} \Rightarrow \eta =
  \frac{\rho_h}{\rho} = \gamma
  \]
  or 
  \[\frac{\rho_t}{\rho_h}=-\frac{R}{c_p}\]
  whose substitution into (B10) leads to the known OB expression for $\omega^2_{bB}$ \cite{3}:
\begin{equation}
  \omega_{bB^{}}^2 = g \left( \gamma_{{ad}} + \gamma_e \right)
\end{equation}
\end{enumerate}

\appendix{}
\newpage
\begin{figure}
\advance\leftskip-3cm
  \includegraphics[scale=0.6]{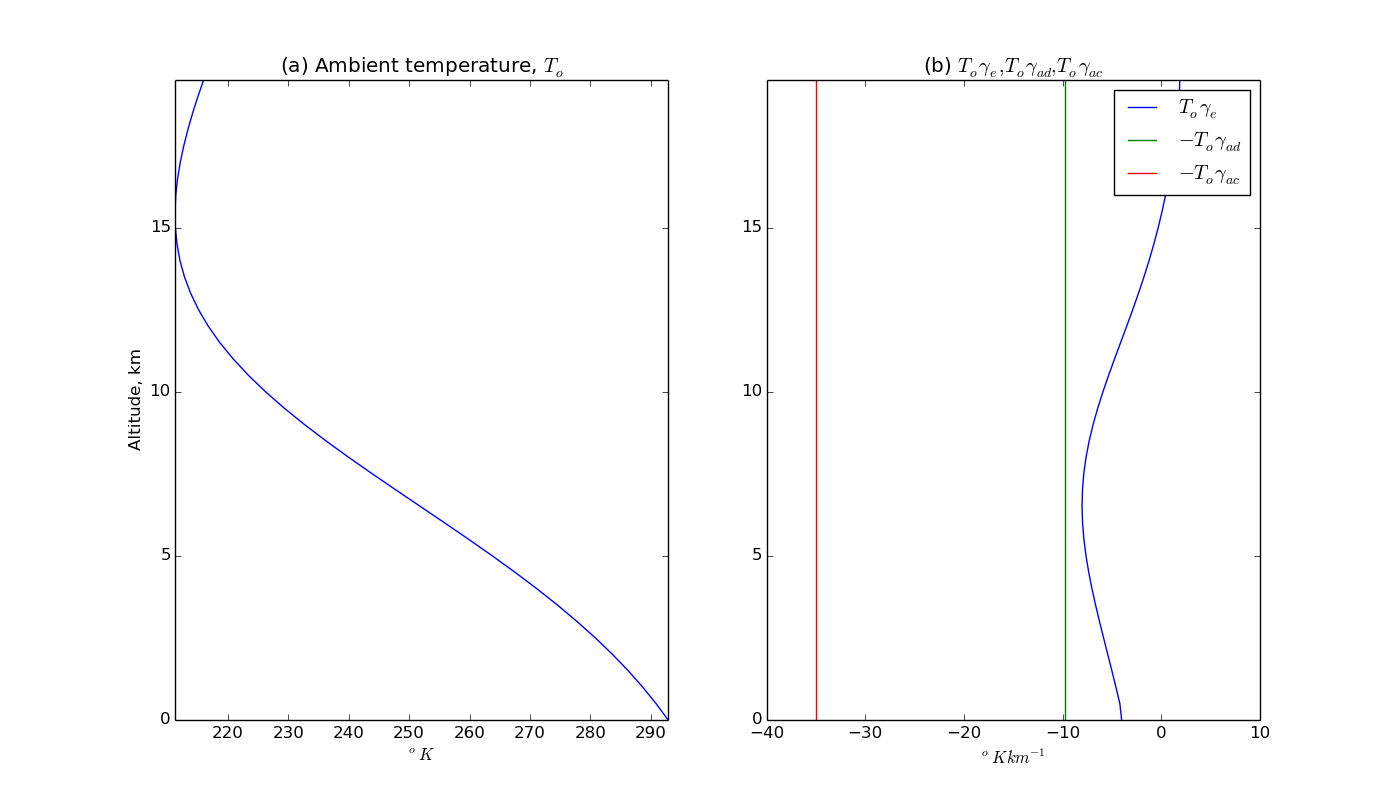}
  \caption{(a) The ambient radiative-convective equilibrium temperature $\left( T_o \right)$
profile, (b) The environmental lapse rate $\gamma_e$ multiplied by $T_o$.
For comparison, the adiabatic and auto-convective lapse rates $\left(
\gamma_{{ad}} \nocomma, \gamma_{{ac}} \right)$, multiplied by $T_o$
are also plotted.}
\end{figure}

\begin{figure}
\advance\leftskip-3cm
  \includegraphics[height=18cm,width=20cm]{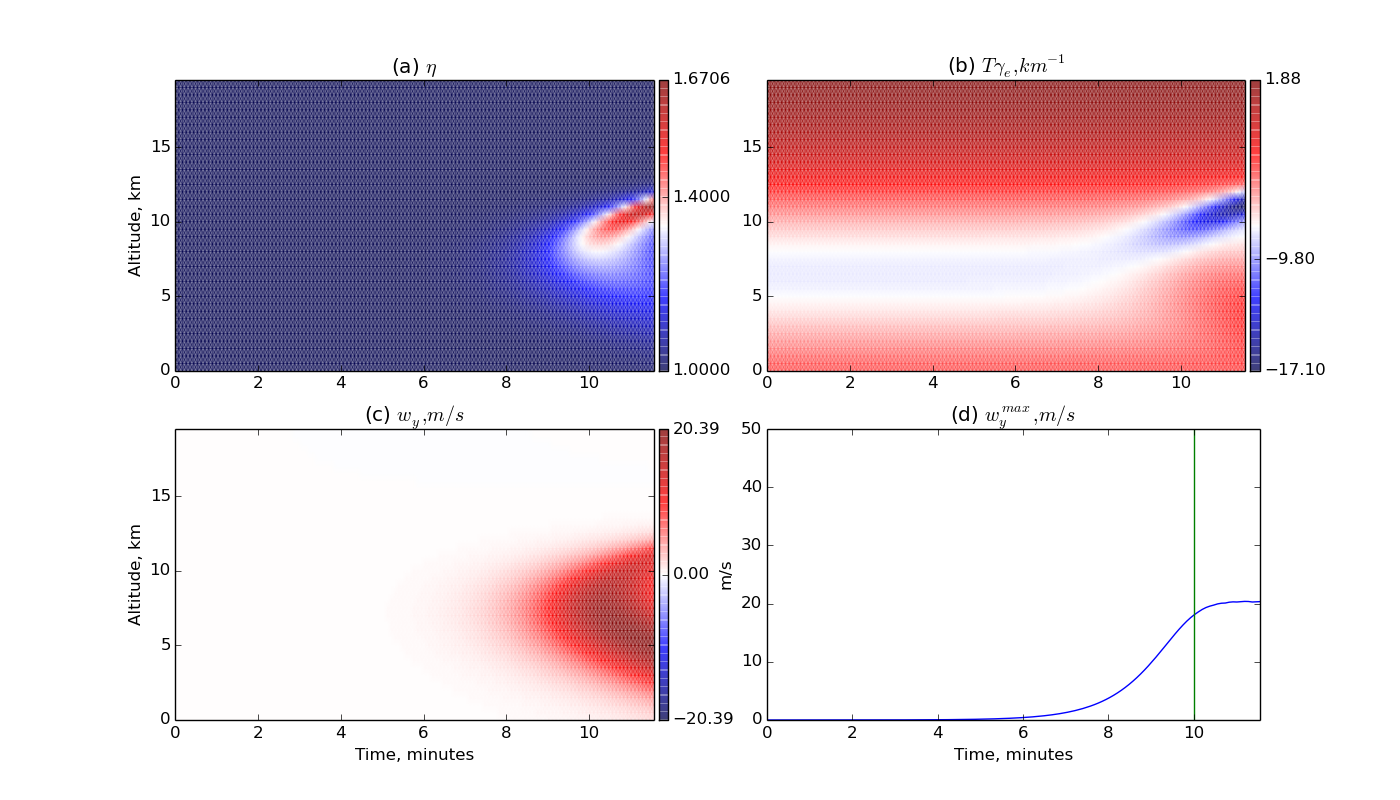}
  \caption{Simulation of CI using governing equations (2) with a wind source: The temporal-altitude
evolution of the (a) $\eta = \frac{\rho_h}{\rho_h + \rho_t} \nocomma$, (b) $T
\gamma_e$, (c) the wind $w_y$ are shown respectively. Here $\left( \rho_h,
\rho_t \right)$ are the hydrostatic equilibrium and perturbed densities,
respectively. In (d) the temporal evolution of the maximum (over altitude) value
$\left( w^{\max}_y \right)$ of $w_y$ is plotted. 
The green line is drawn at the time when $\eta$ approaches $\gamma$.}
\end{figure}

\begin{figure}
\advance\leftskip-3cm
  \includegraphics[height=18cm,width=20cm]{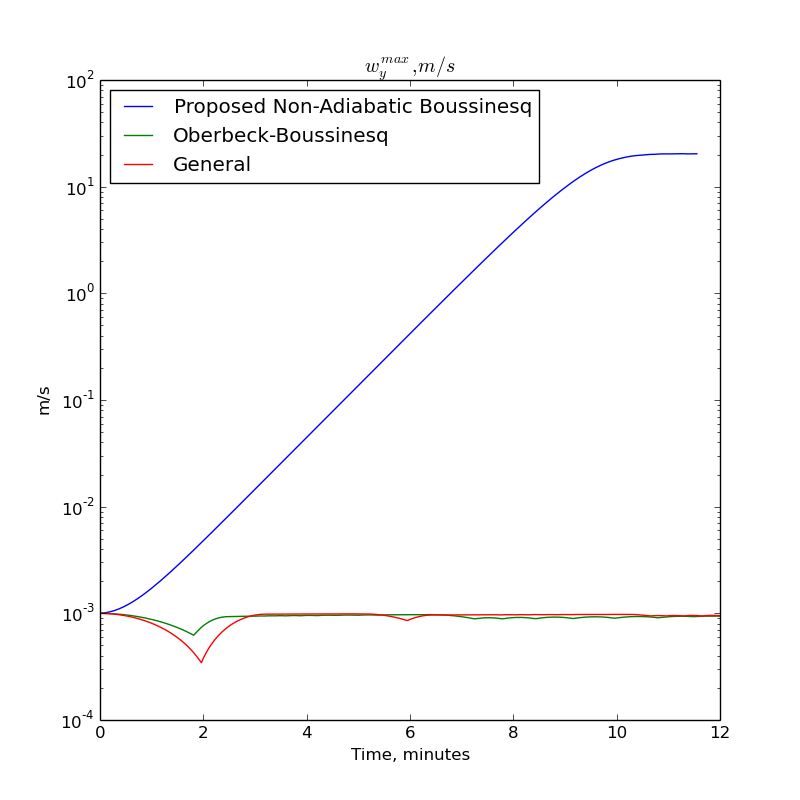}
  \caption{Comparative study from three different approaches: 
  the temporal evolution of the maximum (over altitude) value
$\left( w^{\max}_y \right)$ of $w_y$ is plotted for NAB (blue), OB (green) and the general approach (red).}
\end{figure}

\end{document}